\documentclass[preprint,aps,prd,showpacs,preprintnumbers,amsmath,amssymb,amsfonts,nofootinbib,a4paper,12pt]{revtex4-1}

\usepackage{graphicx}
\usepackage[usenames]{color}
\usepackage{hyperref}

\numberwithin{equation}{section}

\def\stamp{--- {\bf \today} --- {\bf \jobname.tex}}

\def\tr{\textrm{tr}}

\def\cT{\mathcal{T}}

\def\cN{\mathcal{N}}

\def\cZ{\mathcal{Z}}

\def\ZZ{\mathbb{Z}}

\def\stamp{--- {\bf \today} --- {\bf \jobname.tex}}

\def\Ree{\Re\textrm{e}}
\def\Imm{\Im\textrm{m}}

\def\tr{\textrm{tr}}

\def\tr{\textrm{tr}}

\def\cF{\mathcal{F}}
\def\nn{\nonumber}
\def\sg(#1){\textrm{sign}(#1)}

\def\cN{\mathcal{N}}

\def\ZZ{\mathbb{Z}}

\def\an[#1,#2]{\left\langle#1\,#2\right\rangle}
\def\aq[#1,#2,#3]{\left\langle#1|#2|#3\right]}
\def\qa[#1,#2,#3]{\left[#1|#2|#3\right\rangle}
\def\sq[#1,#2]{\left[#1\,#2\right]}
\def\spa#1.#2{\left\langle#1\,#2\right\rangle}
\def\spab#1.#2.#3{\left\langle#1|#2|#3\right]}
\def\spb#1.#2{\left[#1\,#2\right]}
\def\lor#1.#2{\left(#1\,#2\right)}

\begin{document}
%%%%%%%%%%%%%%%%%%%%%%%%%%%%%%%%%%%%%%%%%%%%%%%%%%%%%%%%%%%%%%%%%%%%%%%%%%%%%%%
\preprint{IHES/P/12/02, IPHT-t11/189}
\title{An $R^4$ non-renormalisation theorem in $\mathcal N=4$ supergravity}
\author{{\bf Piotr Tourkine}${}^a$ and {\bf Pierre Vanhove}$^{a,b}$}
\affiliation{ ${}^a$ Institut de Physique Th{\'e}orique, CEA/Saclay, F-91191 Gif-sur-Yvette,
 France\\
${}^b$ IHES,  Le  Bois-Marie, 35  route de  Chartres,
F-91440 Bures-sur-Yvette, France\\
{\tt email: piotr.tourkine,pierre.vanhove@cea.fr}}

\begin{abstract} 
  We consider the four-graviton amplitudes in CHL constructions
  providing four-dimensional $\cN=4$ models with various numbers of vector
  multiplets.  We show that in these models the two-loop amplitude
  has a prefactor of $\partial^2 R^4$.  This implies a non-renormalisation
  theorem for the $R^4$ term, which forbids the appearance of a
  three-loop ultraviolet divergence in four dimensions in the
  four-graviton amplitude.  We connect the special nature of the $R^4$
  term to the $U(1)$ anomaly of pure $\cN=4$ supergravity.
%\stamp
\end{abstract}
\pacs{}

\maketitle

%%%%%%%%%%%%%%%%%%%%%%%%%%%%%%%%%%%%%%%%%%%%%%%%%%%%%%%%%%%%%%%%%%%%%%%%%%%%%
\section{Introduction}\label{sec:introduction}

$\cN=4$ supergravity in four dimensions has sixteen real supercharges
and  $SU(4)$ for R-symmetry group.
The 
gravity supermutiplet  is  composed of a  spin 2 graviton and two spin 0
real scalars in the singlet representation of $SU(4)$, four spin
3/2 gravitini and four spin 1/2 fermions in the fundamental
representation {\bf 4} of $SU(4)$, and six spin 1 gravi-photons in the {\bf 6} of $SU(4)$.
The only matter multiplet is the vector multiplet composed of one spin 1
vector which is $SU(4)$ singlet, four spin 1/2 fermions transforming in
the fundamental of $SU(4)$, and six spin 0 real scalars transforming in
the {\bf 6} of $SU(4)$. The vector multiplets may be carrying
non-Abelian gauge group from a $\cN=4$ super-Yang-Mills theory. 

Pure $\cN=4$ supergravity contains only the gravity supermultiplet and
the two real scalars can be assembled into a complex axion-dilaton scalar $S$
parametrizing the coset space 
$SU(1,1)/U(1)$. This multiplet can be coupled to $n_v$
vector multiplets, whose
scalar fields parametrize the coset space
$SO(6,n_v)/SO(6)\times SO(n_v)$~\cite{de Roo:1984gd}.

$\cN=4$ supergravity theories can be obtained by
consistent dimensional reduction of $\cN=1$ supergravity in $D=10$,
or from various string theory models.
For instance the reduction of the $\cN=8$ gravity super-multiplet
leads to $\cN=4$ gravity super-multiplet, four spin 3/2
$\cN=4$ super-multiplets, and six vector multiplets
\begin{eqnarray}\label{e:multiplet}
  (2_{\bf 1}, 3/2_{\bf 8},1_{\bf 28} , 1/2_{\bf 56},0_{\bf 70})_{\cN=8}&=&   
 (2_{\bf 1}, 3/2_{\bf 4},1_{\bf 6} ,1/2_{\bf 4},0_{\bf 1+1})_{\cN=4}\\
\nn&\oplus&\,{\bf 4}\,
 (3/2_{\bf 1},1_{\bf 4}, 1/2_{\bf 6+1},0_{\bf 4+\bar 4})_{\cN=4}\cr
&\oplus& {\bf 6}\,
 (1_{\bf 1},1/2_{\bf 4},0_{\bf 6})_{\cN=4}\,.
\end{eqnarray}
Removing the four spin 3/2 $\cN=4$ supermultiplets leads to $\cN=4$
supergravity coupled to $n_v=6$ vector multiplets.

 In order to disentangle the contributions from the vector multiplets 
and the gravity supermultiplets, we will use CHL models~\cite{Chaudhuri:1995fk,Chaudhuri:1995bf,Chaudhuri:1995dj}
that allow to construct $\cN=4$ four dimensional
heterotic string with gauge groups of reduced rank. In this
paper we work at a generic point of the
moduli space in the presence of (diagonal) Wilson lines where the
gauge group is Abelian.

Various CHL compactifications in four dimensions can obtained by
considering
 $\ZZ_N$ orbifold~\cite{Chaudhuri:1995bf,Schwarz:1995bj,Aspinwall:1995fw} 
 of the heterotic string  on $T^5\times S^1$. The orbifold
acts  on the current algebra and the right-moving  compactified modes of the
 string
 (world-sheet supersymmetry is on the left moving sector) together with an order $N$ shift  along the $S^1$ direction.
This leads to four-dimensional $\cN=4$ models  with $n_v=48/(N+1)-2$ vector multiplets at a generic
point of the moduli space. Models with $(n_v,N)\in \{(22,1), (14,2), (10,3),
(6,5), (4,7)\}$ have been constructed.  No no-go theorem
  are known ruling out the $n_v=0$ case although it will probably  not
  arise from an asymmetric orbifold construction.\footnote{We would like to thank A. Sen for
a discussion on this point.}

It was shown in~\cite{Bachas:1996bp,Bachas:1997mc,Bachas:1997xn} that $t_8\tr(R^4)$ and
$t_8\tr(R^2)^2$ are half-BPS statured couplings of the heterotic string, receiving
contributions only from the short multiplet of the $\cN=4$ super-algebra, with no
perturbative corrections beyond one-loop. These non-renormalisation theorems were
confirmed in~\cite{D'Hoker:2005ht} using the explicit evaluation of the genus-two
four-graviton heterotic amplitude derived
in~\cite{D'Hoker:2001nj,D'Hoker:2002gw,D'Hoker:2005jc}. For the CHL models, the following fact
is crucially important: the orbifold action does not alter the left moving supersymmetric
sector of the theory. Hence, the fermionic zero mode saturation will happen in the same
manner as it does for the toroidally compactified heterotic string, as
we show in this paper.

Therefore we prove that the genus-two four-graviton amplitude in CHL models satisfy the
same non-renormalisation theorems, due to the factorization at the integrand level of the
mass dimension ten $\partial^2 R^4$ operator in each kinematic
channel. By taking the
field theory limit of this amplitude in four dimensions, no reduction of derivative is
found for generic numbers of vector multiplets $n_v$. Since this result is independent of
$n_v$, we conclude that this rules out the appearance of a $R^4$ ultraviolet counter-term
at three-loop order in four dimensional pure $\cN=4$ supergravity as well. Consequently, the
four-graviton scattering amplitude is ultraviolet finite at three loops in four
dimensions.

The paper is organized as follows. In section~\ref{sec:amplitudes-chl-models} we give the
form of the one- and two-loop four-graviton amplitude in orbifold CHL models.  Then,
in section~\ref{sec:qftlimit} we evaluate their field theory limit in four dimensions.
This gives us the scattering amplitude of four gravitons in $\cN=4$ supergravity coupled
to $n_v$ vector multiplets. In section~\ref{sec:nonrenormalisation} we discuss the
implication of these results for the ultraviolet properties of pure $\cN=4$ supergravity.

\noindent{\bf Note:} As this paper was being finalized, the preprint~\cite{Bern3loopN4} appeared on
the arXiv. In this work  the absence of
three-loop divergence in the four-graviton amplitude in four
dimensions is obtained by a direct field theory computation.

%-------------------------------------------------------------------------------------
\section{One- and Two-loop amplitudes in CHL models}
\label{sec:amplitudes-chl-models}

Our conventions are that the left-moving sector of the heterotic string
is the supersymmetric sector, while the right-moving contains the
current algebra.

We evaluate the four-graviton amplitude in four dimensional CHL
heterotic string models.  We show that the fermionic zero mode
saturation is  model independent and similar to the
toroidal compactification.

%-------------------------------------------------------------------------------------
\subsection{The one-loop amplitude in string theory}
\label{sec:one-loop-string}

The expression of the one-loop four-graviton amplitude in CHL models
in $D=10-d$ dimensions is an immediate extension of the amplitude
derived in~\cite{Sakai:1986bi}

\begin{equation}
  \label{e:4gravHet}
 \mathcal M^{(n_v)}_{4,1-loop} =\cN_1\, \int_{\cF} {d^2\tau\over\tau_2^{2-{d\over2}}}
  \cZ^{(n_v)}_1\, \int_{\cT}
  \prod_{1\leq i<j\leq 4} {d^2\nu_i\over\tau_2}\,\mathcal W^{(1)} \,
  e^{-\sum_{1\leq i<j\leq4}2\alpha' k_i\cdot k_j P(\nu_{ij})}\,,
\end{equation}
where $\cN_1$ 
is a constant of normalisation, $\cF:=\{\tau=\tau_1+i\tau_2, |\tau|\geq1, |\tau_1|\leq
\frac12, \tau_2>0\}$ is a fundamental domain for $SL(2,\ZZ)$ and
the domain of integration $\cT$ is defined as
$\cT:=\{\nu=\nu^1+i\nu^2; |\nu^1|\leq\frac12, 0\leq \nu^2\leq\tau_2\}$.
$\cZ^{(n_v)}_1$ is the genus-one partition function of the CHL model.

The polarisation of the $r$th graviton is factorized as
$h^{(r)}_{\mu\nu}=\epsilon^{(r)}_\mu\,\tilde\epsilon^{(r)}_\nu$. We introduce the notation $t_8F^4:= t_8^{\mu_1\cdots \mu_8} \prod_{r=1}^4 k^{(r)}_{\mu_{2r-1}}\, 
  \epsilon^{(r)}_{\mu_{2r}}$.
The quantity $\mathcal W^{(1)}$ arises from the contractions of the
right-moving part of the graviton vertex operator

\begin{equation}
  \label{e:W1}
  \mathcal W^{(1)}:=t_8F^4\,{\langle \prod_{j=1}^4\tilde \epsilon^j \cdot \bar\partial X(z_j)
    e^{i k_j\cdot x(z_j)} \rangle\over \langle  \prod_{j=1}^4 
    e^{i k_j\cdot x(z_j)}  \rangle} 
=t_8F^4\,\prod_{r=1}^4\tilde\epsilon^{(r)}_{\nu_r}\, t_{4;1}^{\nu_1\cdots\nu_4}\,,
\end{equation}
with  $\hat t_{4;1}^{\nu_1\cdots \nu_4}$ the quantity evaluated in~\cite{Sakai:1986bi}

\begin{equation}
  \label{e:thatone}
\hat  t_{4;1}^{\nu_1\cdots\nu_4}:= Q_1^{\nu_1}\cdots Q_4^{\nu_4} + {1\over2\alpha'}
  \,(Q_1^{\nu_1}Q_2^{\nu_2}\delta^{\nu_3\nu_4}
  T(\nu_{34})+perms)+{1\over4{\alpha'}^2}\,
  (\delta^{\nu_1\nu_2}\delta^{\nu_3\nu_4} T(\nu_{12})T(\nu_{34})+perms)\,,
\end{equation}
where
\begin{equation}\label{e:QT}
  Q_I^\mu:= \sum_{r=1}^4 \, k^{(r)\mu}\, \bar\partial
  P(\nu_{Ir}|\tau);\qquad 
 T(\nu):= \bar\partial_\nu^2 P(\nu|\tau)\,.
\end{equation}
We follow the notations and conventions
of~\cite{Green:1999pv,Green:2008uj}. The genus one propagator is given by

\begin{equation}
  \label{e:Prop}
  P(\nu|\tau):= -\frac14\,
  \log\left|\theta_1(\nu|\tau)\over \theta_1'(0|\tau) \right|^2+ {\pi
    (\nu^2)^2\over2\tau_2}\,.
\end{equation}

In the  $\alpha'\to0$ limit relevant for the field theory analysis in
section~\ref{sec:qftlimit}, with all the radii of compactification 
scaling like $\sqrt{\alpha'}$,
the mass of the Kaluza-Klein excitations and winding modes go to
infinity and the genus-one partition function  $\cZ_1^{(n_v)}$ has the
following  expansion in $\bar q=\exp(-2i\pi \bar\tau)$
\begin{equation}
  \label{e:cLnv}
  \mathcal \cZ^{(n_v)}_1= {1\over \bar q}+ c_{n_v}^1  +O(\bar q)\,.
\end{equation}
The $1/\bar q$ contribution is the ``tachyonic'' pole, $c^1_{n_v}$ depends on the
number of vector multiplets and higher orders in $\bar q$ coming from to massive
string states do not contribute in the field theory limit. 

%-------------------------------------------------------------------------------------
\subsection{The two-loop amplitude in string theory}
\label{sec:two-loop-string}

By applying the techniques for evaluating  heterotic string
two-loop amplitudes  of~\cite{D'Hoker:2001nj,D'Hoker:2002gw,D'Hoker:2005ht,D'Hoker:2005jc},
we obtain that the four-graviton amplitudes in the CHL models are given by 
\begin{equation}\label{e:twoloopstring}
\mathcal  M^{(n_v)}_{4,2-loop}= \mathcal N_2\, \int {|d^3\Omega|^2\over
    (\det\Imm\Omega)^{5-{d\over2}}} \,\cZ^{(n_v)}_2\,\int \prod_{i=1}^4 d^2\nu_i\, \mathcal W^{(2)}\, {\mathcal Y}_s \, e^{-\sum_{1\leq i<j\leq 4}2\alpha'
    k^i\cdot k^j P(\nu_{ij})}
\end{equation}
where $\mathcal N_2$ is a normalization constant, 
 $\cZ_2^{(n_v)}(\Omega,\bar\Omega)$ is the genus-two partition function and 
\begin{equation}\label{e:W2genus2}
  \mathcal W^{(2)}:=t_8F^4\, {\langle \prod_{j=1}^4 \epsilon^j \cdot \bar\partial X(z_j)
    e^{i k_j\cdot x(z_j)} \rangle\over \langle  \prod_{j=1}^4 
    e^{i k_j\cdot x(z_j)}  \rangle}=t_8F^4\, \prod_{i=1}^4 \tilde \epsilon_i^{\nu_i}\, t_{4;2}^{\nu_1\cdot\nu_4} \,.
\end{equation}
The tensor  $ t_{4;2}^{\nu_1\cdot\nu_4}$ is the genus-two equivalent of the
genus-one tensor given in~\eqref{e:thatone}
\begin{equation}
  \label{e:t4hattwo}
  t_{4;2}^{\nu_1\cdots\nu_4}=
 Q_1^{\nu_1}\cdots Q_4^{\nu_4}+{1\over2\alpha'} Q_1^{\nu_1}Q_2^{\nu_2}T(\nu_{34})\delta^{\nu_3\nu_4}+{1\over4(\alpha')^2}\,\delta^{\nu_1\nu_2}\delta^{\nu_3\nu_4}T(\nu_{12})T(\nu_{34})+perms\,,
\end{equation}
 this time expressed in terms of the genus-two bosonic propagator

\begin{equation}
  \label{e:P2}
  P(\nu_1-\nu_2|\Omega):=-\log|E(\nu_1,\nu_2|\Omega)|^2+2\pi
  (\Imm\Omega)_{IJ}^{-1}\,(\Imm\int_{\nu_1}^{\nu_2} \omega_I)(\Imm\int_{\nu_1}^{\nu_2} \omega_J)\,,
\end{equation}
where  $E(\nu)$ is the genus-two prime form, $\Omega$ is the period
matrix and $\omega_I$ with $I=1,2$ are the holomorphic abelian differentials. We refer to~\cite[Appendix~A]{D'Hoker:2005jc} for the
main properties of these objects.

The $\mathcal Y_S$  quantity, arising from several contributions in the
RNS formalism  and from the fermionic zero modes in the pure spinor formalism~\cite{Berkovits:2005df,Berkovits:2005ng},  is given by 
\begin{equation}\label{e:Ys}
3  \mathcal Y_S= (k_1-k_2)\cdot (k_3-k_4)\, \Delta_{12}\Delta_{34}+
(13)(24) + (14)(23)\,,
\end{equation}
with
\begin{equation}
  \Delta(z,w)=\omega_1  (z) \omega_2(w)-\omega_1(w) \omega_2(z)\,.
\end{equation}
Using the identity $\Delta_{12} \Delta_{34}+\Delta_{13}
\Delta_{42}+\Delta_{14} \Delta_{23}=0$ we have the
equivalent form $\mathcal Y_S=-3\,(s
\Delta_{14}\Delta_{23}-t\Delta_{12}\Delta_{34})$, where
$s=(k_1+k_2)^2$, $t=(k_1+k_4)^2$ and $u=(k_1+k_3)^2$.

We use a parametrisation of the period matrix reflecting the symmetries of
the field theory vacuum two-loop diagram considered in the next section 

\begin{equation}
  \label{e:OmegaBis}
  \Omega:=
  \begin{pmatrix}
   \tau_1+\tau_3 & \tau_3 \cr \tau_3 & \tau_2+\tau_3
  \end{pmatrix}
\,.
\end{equation}
With this parametrisation the expression for $\cZ^{(n_v)}_2(\Omega,\bar\Omega)$ is completely symmetric in  the
variables $q_I=\exp(2i\pi \tau_I)$ with $I=1,2,3$. 

 In the limit relevant for the field theory analysis in
section~\ref{sec:qftlimit}, 
the partition function of the CHL model
has the following $\bar q_i$-expansion~\cite{TourkineVanhove}

\begin{equation}
  \label{e:cl2limit}
  \cZ^{(n_v)}_2= {1\over \bar q_1\bar q_2\bar q_3}+ a_{n_v}
  \sum_{1\leq i<j\leq3} {1\over \bar q_i\bar q_j}+ b_{n_v} \sum_{1\leq i\leq3} {1\over \bar q_i}+c_{n_v}+O(q_i)\,.
\end{equation}

%%%%%%%%%%%%%%%%%%%%%%%%%%%%%%%%%%%%%%%%%%%%%%%%%%%%%%%%%%%%%%%%%%%%%%%%%%%%%%%%%%%%%%
\section{The field theory limit}\label{sec:qftlimit}

In this section we extract the field theory
limit of the string theory amplitudes compactified to four dimensions.
We consider the low-energy limit $\alpha'\to0$ with the radii
of the torus scaling like $\sqrt{\alpha'}$ so that all the massive
Kaluza-Klein states, winding states and excited string states
decouple.

In order to simplify the analysis we make the following choice of
polarisations
$(1^{++},2^{++},3^{--},4^{--})$ and of reference momenta\footnote{Our conventions
  are that a null vector $k^2=0$ is parametrized by $k_{\alpha\dot\alpha}=
k_{\alpha} \bar k_{\dot\alpha}$. 
The spin 1  polarisations of positive and negative helicities are
given by 
$
  \epsilon^+(k,q)_{\alpha\dot\alpha}:= {q_\alpha \bar
    k_{\dot\alpha}\over\sqrt2\, \an[q,k]}$, 
$ \epsilon^-(k,q)_{\alpha\dot\alpha}:=- {k_\alpha \bar
    q_{\dot\alpha}\over\sqrt2\, \sq[q,k]}$,
where $q$ is a reference momentum.
One finds that $
t_8F^{(1)+}\cdots F^{(4)+}= 
t_8F^{(1)-}\cdots F^{(4)-}=0$ and $t_8F^{(1)-}F^{(2)-}F^{(3)+} F^{(4)+}={1\over16}
{\an[k_1,k_2]^2\sq[k_3,k_4]^2}$}
$q_1=q_2=k_3$ and $q_3=q_4=k_1$, such that
$2t_8F^4=\an[k_1,k_2]^2\sq[k_3,k_4]^2$, and
$4t_8t_8R^4=\an[k_1,k_2]^4\sq[k_3,k_4]^4$.
With these choices the expression for $\mathcal W^{(g)}$ reduces to 

\begin{eqnarray}
\nn 
  \mathcal W^{(g)}&=& t_8t_8R^4\, 
    (\bar\partial P(\nu_{12})-\bar\partial P(\nu_{14}))
    (\bar\partial P(\nu_{21})-\bar\partial P(\nu_{24}))
    (\bar\partial P(\nu_{32})-\bar\partial P(\nu_{34}))
    (\bar\partial P(\nu_{42})-\bar\partial P(\nu_{43}))\\
&+& {t_8t_8R^4\over u}\, \bar\partial^2 P(\nu_{24}) 
 (\bar\partial P(\nu_{12})-\bar\partial P(\nu_{14}))
 (\bar\partial P(\nu_{32})-\bar\partial P(\nu_{34}))\,,
 \label{e:Wg}\end{eqnarray}
where $s=(k_1+k_2)^2$, $t=(k_1+k_4)^2$ and $u=(k_1+k_3)^2$.
We introduce the notation $\mathcal W^{(g)}= t_8t_8R^4\, (\mathcal
W^{(g)}_1+ u^{-1} \mathcal W^{(g)}_2)$.

The main result of this section is
that the one-loop amplitudes factorizes a $t_8t_8R^4$ and that the
two-loop amplitudes factorizes a $\partial^2 t_8t_8 R^4$ term.
A more detailed analysis  will be given in the work~\cite{TourkineVanhove}.

%-------------------------------------------------------------------------------
\subsection{The one-loop amplitude in field theory}
\label{sec:one-loop-ft}

In the field theory limit $\alpha'\to0$ and $\tau_2\to\infty$ with
$t=\alpha'\tau_2$  fixed, we define $\nu^2=\tau_2\,
\omega$ for $\nu=\nu^1+i\nu^2$.

Because of the $1/\bar q$ pole in the partition
function~\eqref{e:cLnv} the integration over $\tau_1$ yields two contributions
\begin{equation}
\int_{-\frac12}^{\frac12} d\tau_1 \,\cZ_1^{(n_v)} F(\tau,\bar \tau)
= F_1+ c_{n_v}^1\, F_0\,,
\end{equation}
where $F(\tau,\bar \tau)= F_0+ \bar q F_1 + c.c. +O(\bar q^2)$ represents the
integrand of the one-loop amplitude.

The bosonic propagator can be split in an asymptotic value for $\tau_{2}\to\infty$
(the field theory limit) and a correction~\cite{Green:1999pv}

\begin{equation}\label{e:Pasymp1} 
P(\nu|\tau)
=P^\infty(\nu|\tau)+\hat P(\nu|\tau)
\end{equation}
that write:
\begin{eqnarray}
P^\infty(\nu|\tau)&=&{\pi (\nu^2)^2\over 2\tau_2}- {1\over4} \ln\left|\sin(\pi \nu)\over \pi \right|^2 \cr
\hat P(\nu|\tau)&=& -
\sum_{m\geq1} \left({q^m\over 1-q^m} {\sin^2(m\pi \nu)\over m}
  +c.c.\right)+C(\tau),
\end{eqnarray}
  where $q= \exp(2i\pi\tau)$ and $C(\tau)$ is a zero mode contribution which drops out of the amplitude due to the momentum 
 conservation \cite{Green:1999pv}.

We decompose the asymptotic propagator
$P^\infty(\nu|\tau)={\pi\over2}\,\tau_2\, P^{FT}(\omega)+\delta_s(\nu)$
into  a piece that will
dominate in the field theory limit

\begin{equation}
  \label{e:PFT}
  P^{FT}(\omega) = \omega^2 - |\omega|\,,
\end{equation}
 and  a  contribution $\delta_s(\nu)$  from the massive string modes~\cite[appendix~A]{Green:1999pv}
 \begin{equation}
   \label{e:delta}
   \delta_s(\nu):= \sum_{m\neq0} {1\over 4|m|} \, e^{2i\pi m
       \nu^1-2\pi |m\nu^2|}\,.
 \end{equation}

The expression for $Q_I^\mu$ and  $T$ in~\eqref{e:QT} become
\begin{eqnarray}
  \label{eq:3}
  Q_I^\mu &= &Q_I^{FT\, \mu}+ \delta Q_I^\mu -\pi \sum_{r=1}^4 k^{(r)\mu}\,
  \sin(2\pi \bar\nu_{Ir}) \, \bar q+o(\bar q^2)\\
\nn  T(\bar \nu)&=&T^{FT}(\omega)+\delta T(\bar \nu)+ 2\pi \cos(2\pi \bar \nu)\,
\bar q+o(\bar q^2)\,,
\end{eqnarray}
where
\begin{eqnarray}
  \label{e:QTFT}
  Q^{FT\,\mu}_I&:=&-{\pi\over2}\, (2K^\mu+q^{\mu}_I)\\
\label{e:K} K^\mu&:=& \sum_{r=1}^4 \, k^{(r)\mu}\, \omega_r\\
\label{e:qr} q^{\mu}_I&:=&\sum_{r=1}^4
  k^{(r)\mu}\, \sg(\omega_I-\omega_r)\\
T^{FT}(\omega)&=&{\pi\alpha'\over t} (1-\delta(\omega))\,,
\end{eqnarray}
and 

\begin{eqnarray}
  \label{e:deltaQT}
  \delta Q_I^\mu(\bar \nu)&=&\sum_{r=1}^4 k^{(r)\mu}\, \bar\partial
  \delta_s(\bar\nu_{Ir})
=-\frac{i\pi}2\sum_{r=1}^4 \, \sg (\nu^2_{Ir})k^{(r)\mu} \sum_{m\geq1} e^{-\sg (\nu^2_{Ir})\,2i\pi
    m\bar\nu_{Ir}}\\
\nn \delta T(\nu)&=&\bar\partial^2 \delta_s(\bar\nu)
=-\pi^2 \, \sum_{m\geq1} m \,  e^{-\sg (\nu^2_{Ir})\,2i\pi
    m\bar\nu_{Ir}}\,.
\end{eqnarray}

We introduce the notation 
\begin{equation}
  \label{e:Q}
  Q^{(1)}(\omega):=\sum_{1\leq i<j\leq 4} k_i\cdot k_j \, P^{FT}(\omega_{ij})\,,
\end{equation}
such that  $\partial_{\omega_i} Q^{(1)}=  k_i\cdot Q_i^{FT}$.

In the field theory limit $\alpha'\to0$ the integrand of the string
amplitude in~\eqref{e:4gravHet} becomes 
\begin{eqnarray}
  M_{4;1}^{(n_v)}&=& N_1\,t_8t_8R^4\,\int^\infty_0 {d\tau_2\over
    \tau_2^{2-{d\over2}}} \int_{\Delta_\omega}
  \prod_{i=1}^3 d\omega_i\,\, e^{t\,Q^{(1)}(\omega)}\times\\
\nn&\times&\int_{-\frac12}^{\frac12}d\tau_1 \int_{-\frac12}^{\frac12}
  \prod_{i=1}^4 d\nu^1_i\, {1+c^1_{n_v} \bar q+o(\bar q^2)\over\bar
    q}\,(\mathcal W^{(1)}_1+ {1\over u} \mathcal W^{(1)}_2)\,\times\\
\nn&\times& \exp\left(\sum_{1\leq i<j\leq
    4}2\alpha'\,k_i\cdot k_j \,\left(\delta_s(\nu_{ij})-\sum_{m\geq1} \bar q\,\sin^2(\pi
    \bar\nu_{ij}) +O(\bar q)\right)\right)\,,
\end{eqnarray}
here $N_1$ is a constant of normalisation. The
 domain of integration $\Delta_\omega=[0,1]^3$ is decomposed into
three regions $\Delta_w=\Delta_{(s,t)}\cup \Delta_{(s,u)}\cup \Delta_{(t,u)} $  given by the union of the
$(s,t)$, $(s,u)$ and $(t,u)$ domains. In the $\Delta_{(s,t)}$ domain the
integration is performed over  $0\leq \omega_1\leq
\omega_2\leq\omega_3\leq1$ where $Q^{(1)}(\omega)=-s\omega_1(\omega_3-\omega_2)-t(\omega_2-\omega_1)(1-\omega_3)$ with equivalent
formulas obtained by permuting the external legs labels in the $(t,u)$ and $(s,u)$ regions (see~\cite{Green:1999pv} for details).

The leading contribution to the amplitude is given by
\begin{eqnarray}
&&  M_{4;1}^{(n_v)}= N_1\,t_8t_8R^4\,\int^\infty_0 {d\tau_2\over
    \tau_2^{2-{d\over2}}} \int_{\Delta_\omega}
  \prod_{i=1}^3 d\omega_i\,\, e^{t\,Q^{(1)}(\omega)}\times\\
\nn&\times&\int_{-\frac12}^{\frac12}
  \prod_{i=1}^4 d\nu^1_{i}\, \left(\left.\left(\mathcal W^{(1)}_1+{1\over
    u}\mathcal W^{(1)}_2\right)\right|_0 (c^1_{n_v}-\sum_{1\leq i<j\leq 4}2\alpha'\,
k_i\cdot k_j\, \sin^2(\pi\bar \nu_{ij}))+ \left.\left(\mathcal W^{(1)}_1+{1\over
    u}\mathcal W^{(1)}_2\right)\right|_1\right)\,,
\end{eqnarray}
where $(\mathcal W^{(1)}_1+{1\over
    u}\mathcal W^{(1)}_2)|_0$ and $(\mathcal W^{(1)}_1+{1\over
    u}\mathcal W^{(1)}_2)|_1$ are respectively the zeroth and first order
in the $\bar q$ expansion of $\mathcal W^{(1)}_i$.

Performing the integrations over the $\nu^1_{i}$ variables leads
to the following structure for the amplitude reflecting the
decomposition in~\eqref{e:multiplet} 

\begin{equation}
  \label{e:oneloophet}
  M_{4;1}^{(n_v)}=N_1{\pi^4\over4}\,\left(c^1_{n_v} \,
    M_{4;1}^{\rm\cN=4~matter}+ M_{4;1}^{\cN=8}- 4 M_{4;1}^{\cN=4~spin~\frac32}
 \right)\,.
\end{equation}
The contribution from the $\cN=8$ supergravity multiplet is given by
the quantity evaluated in~\cite{Green:1982sw}

\begin{equation}\label{e:N=8}
    M^{\cN=8}_{4;1} = t_8t_8R^4\, 
  \int_{\Delta_\omega} d^3\omega\,\Gamma\left(2+\epsilon\right)\, (Q^{(1)})^{-2-\epsilon}\,,
\end{equation}
where we have specified the dimension $D=4-2\epsilon$ and $Q^{(1)}$ is
defined in~\eqref{e:Q}.
The contribution from the $\cN=4$ matter fields vector multiplets is

\begin{equation}\label{e:N=4matter}
    M^{\cN=4~matter}_{4;1} =t_8t_8R^4\,{\pi^4\over16}\,
  \int_{\Delta_\omega} \!\!\!d^3\omega\!\Big[ \Gamma\left(1+\epsilon\right)\,
  (Q^{(1)})^{-1-\epsilon}\, W_2^{(1)}
+\Gamma\left(2+\epsilon\right)\, (Q^{(1)})^{-2-\epsilon} \, W_1^{(1)}\Big]
\end{equation}
where $W_i^{(1)}$ with $i=1,2$ are the field theory limits of the
$\mathcal W_i^{(1)}$'s
\begin{eqnarray}
  \label{e:Tneval}
\nn  W_2^{(1)}&=&{1\over u}\, (2\omega_2-1+\sg(\omega_3-\omega_2))  (2\omega_2-1+\sg(\omega_1-\omega_2))\,(1-\delta(\omega_{24}))\\
\nn W_1^{(1)}&=&2
( \omega_2- \omega_3) (\sg(\omega_1-\omega_2)+2 \omega_2-1)\times\\
&\times&
   (\sg(\omega_2-\omega_1)+2 \omega_1-1) (\sg(\omega_3-\omega_2)+2
   \omega_2-1)\,.
\end{eqnarray}
Finally, the $\cN=6$ spin $3/2$ gravitino multiplet running in the
loop gives
\begin{equation}\label{e:N=4Gravitino1loop}
  M^{\rm \cN=6~spin~\frac32}_{4;1} = t_8 t_8 R^4 \, 
  \int_{\Delta_\omega}
  d^3\omega\,\Gamma\left(2+\epsilon\right)\tilde W_2^{(1)}\, (Q^{(1)})^{-2-\epsilon},
\end{equation}
where 
\begin{eqnarray}
\tilde W_2^{(1)} &=& 
(\sg(\omega_1-\omega_2)+2 \omega_2-1)(\sg(\omega_2-\omega_1)+2 \omega_1-1)\\
\nn&+& (\sg(\omega_3-\omega_2)+2\omega_2-1)( \omega_3- \omega_2) \,.
\end{eqnarray}
The $\cN=6$ spin $3/2$ supermultiplet is the sum of a $\cN=4$ spin
$3/2$ supermultiplet and two $\cN=4$ spin 1 supermultiplet.

Using the dictionary given
in~\cite{BjerrumBohr:2008vc,BjerrumBohr:2008ji}, we recognize that 
the amplitudes in~\eqref{e:N=4matter} and~\eqref{e:N=4Gravitino1loop}
are combinations of  scalar box integral functions $I_4^{(D=4-2\epsilon)}[\ell^n]$ evaluated in
$D=4-2\epsilon$  with $n=4,2,0$ powers of loop momentum and
$I_4^{(D=6-2\epsilon)}[\ell^n]$ with $n=2,0$ powers of loop momentum evaluated in
$D=6-2\epsilon$ dimensions. The $\cN=8$ supergravity part in~\eqref{e:N=8}
 is only given by a scalar box amplitude function $I_4^{(D=4-2\epsilon)}[1]$ evaluated
 in $D=4-2\epsilon$ dimensions.

Those amplitudes are free of ultraviolet divergences but
exhibit rational terms, in agreement with
the analysis of ~\cite{Dunbar:2010fy,Dunbar:2011xw,Dunbar:2011dw,Bern:2011rj}. This was
not obvious from the start, since superficial power counting indicates a logarithmic
divergence. More generally, in $\cN=4$ supergravity models coupled to
vector multiplets amplitudes with external vector multiplets are
ultraviolet divergent at one-loop~\cite{Fischler:1979yk}\footnote{We would like thank K.S.
Stelle and Mike Duff for a discussion about this.}. 

%-------------------------------------------------------------------------------
\subsection{The two-loop amplitude in field theory}
\label{sec:two-loop-ft}
\begin{figure}
  \centering
\includegraphics[width=15cm]{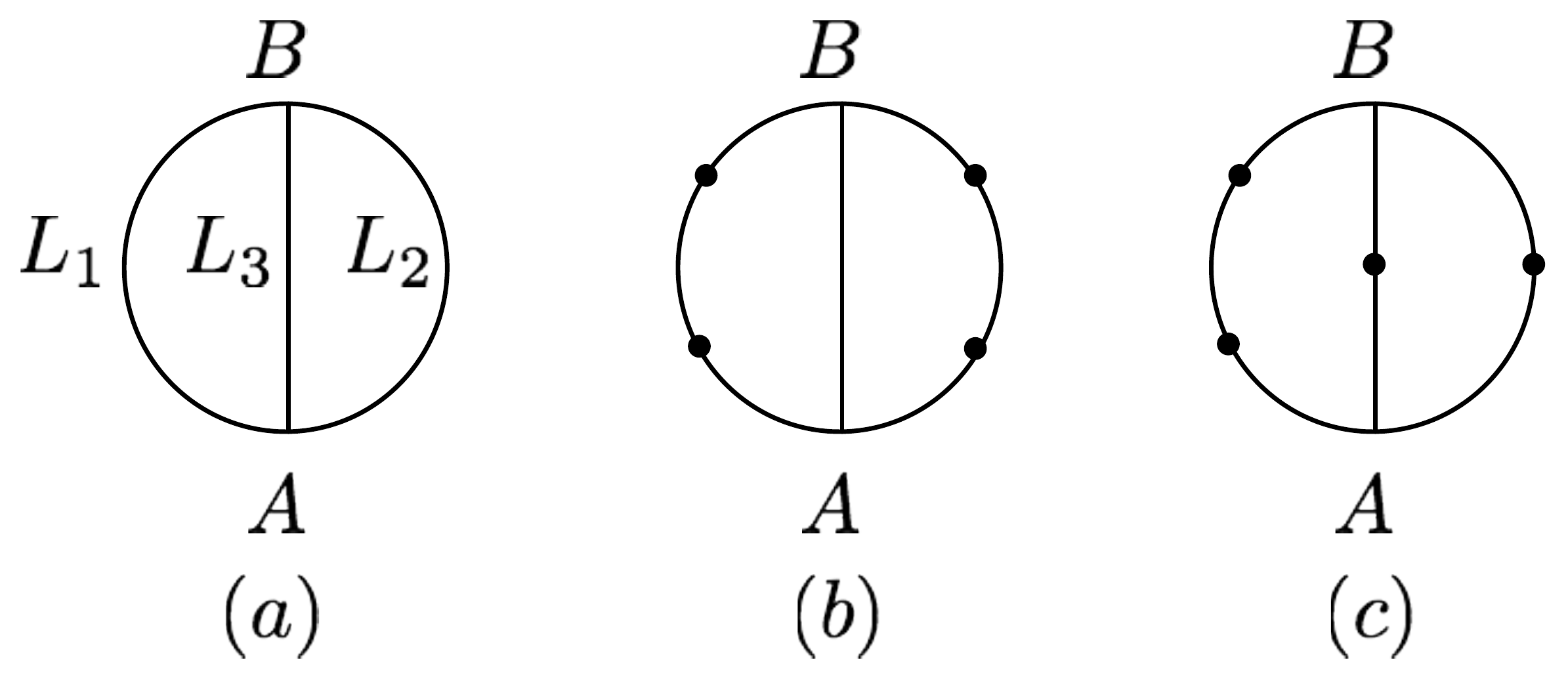}  
  \caption{\sl Parametrisation of the two-loop diagram in field
    theory. Figure~(a) is the vacuum diagram and the definition of the proper times, and
    figures~(b) and~(c) the two configurations contributing to the
    four-point amplitude.}
  \label{fig:twoloop}
\end{figure}

We will follow the notations
of~\cite[section~2.1]{Green:2008bf} where the two-loop four-graviton
amplitude in $\cN=8$ supergravity was presented in the world-line formalism.
In the field theory limit $\alpha'\to0$ the imaginary part of the
genus-two period matrix $\Omega$ becomes the period matrix  $K:=\alpha'\Imm \Omega$
of the two-loop graph in figure~\ref{fig:twoloop}
\begin{equation}
  \label{e:Kdef}
  K:=
  \begin{pmatrix}
    L_1+L_3& L_3\cr L_3& L_2+L_3
  \end{pmatrix}
\,.
\end{equation}
We set $L_i=\alpha'\, \tau_i$ and $\Delta = \det K=L_1L_2+L_1L_3+L_2L_3$.
The position of a point on the line $l=1,2,3$ of length 
$L_l$ will be denoted by $t^{(l)}$. We choose the point $A$ to be the origin of the
coordinate system, i.e.  $t^{(l)}=0$ means the point is located at position $A$,
and $t^{(l)}=L_l$ on the $l$th line means the point is located at position
$B$.

It is convenient to introduce the rank two vectors $v_i= t_i^{(l_i)}\, u^{(l_i)}$ where
\begin{equation}
  \label{e:uI}
  u^{(1)}:=
  \begin{pmatrix}
    1\cr 0
  \end{pmatrix},\quad
 u^{(2)}:= \begin{pmatrix}
    0\cr 1
  \end{pmatrix},\quad
 u^{(3)}:=\begin{pmatrix}
    -1\cr -1
  \end{pmatrix}
 \,.
\end{equation}
The $v_i$ are the  field theory degenerate form of the Abel map
of a point on the Riemann surface to its divisor. The vectors $u^{(i)}$ are the
degenerate form of the integrals of the holomorphic one-forms
$\omega_I$.
If the integrations on each line are oriented from $A$ to $B$, the
integration element on line $i$ is $du^{l_i}= dt_i \, u^{(l_i)}$. The
canonical homology basis $(A_i,B_i)$ of the genus two Riemann surface degenerates
to $(0,b_i)$, with 
 $b_i=L_i \cup \bar L_3$.  $\bar L_3$ means that we
circulate on the middle line from $B$ to $A$. With these definitions we can reconstruct the period
matrix~\eqref{e:Kdef} from 
\begin{eqnarray}
  \label{e:Kint}
 \nn \oint_{b_1} du\cdot u^{(1)}&=&\int_0^{L_1} dt_1 + \int_0^{L_3} dt_3  =L_1+L_3\\
\nn  \oint_{b_2} du\cdot u^{(2)}&=&\int_0^{L_2} dt_1 + \int_0^{L_3} dt_3
  =L_2+L_3\\
\nn  \oint_{b_1} du\cdot u^{(2)}&=& \int_0^{L_3} dt_3
  =L_3\\
\oint_{b_2} du\cdot u^{(1)}&=& \int_0^{L_3} dt_3  =L_3\,,
\end{eqnarray}
in agreement with the corresponding relations on the Riemann surface
$\oint_{B_I} \omega_J= \Omega_{IJ}$.
In the field theory limit, $\mathcal Y_S$ \eqref{e:Ys} becomes
\begin{equation}
\label{e:YsFT}
3Y_S=    (k_1-k_2)\cdot (k_3-k_4)\, \Delta^{FT}_{12}\Delta^{FT}_{34}+
(13)(24) + (14)(23)
\end{equation}
where
\begin{equation}
  \label{e:1DeltaFT}
  \Delta^{FT}_{ij}= \epsilon^{IJ} u_I^{(l_i)} u_J^{(l_j)}\,.
\end{equation}
Notice that $\Delta^{FT}_{ij}=0$ when the point $i$ and $j$ are on the same
line (i.e. $l_i=l_j$). Therefore $Y_S$ vanishes if three points are on
the same line, and the only non-vanishing configurations are the one
depicted in figure~\ref{fig:twoloop}(b)-(c).

In the field theory limit the leading contribution to $Y_S$ is given
by 
\begin{equation}
  \label{e:Ysft}
  Y_S=
  \begin{cases}
    s&\textrm{for}~l_1=l_2~\textrm{or}~l_3=l_4\cr
t&\textrm{for}~l_1=l_4~\textrm{or}~l_3=l_2\cr
u&\textrm{for}~l_1=l_3~\textrm{or}~l_2=l_4\,.\cr
  \end{cases}
\end{equation}
The bosonic propagator in~\eqref{e:P2} becomes

\begin{equation}
  \label{e:Pft2}
  P^{FT}_2(v_i-v_j):=-\frac12\, d(v_i-v_j)+\frac12\, (v_i-v_j)^T \,
  K^{-1}\, (v_i-v_j)\,,
\end{equation}
where $d(v_i-v_j)$ is  given by $|t_i^{(l_i)}-t_j^{(l_j)}|$ if the two
points are on the same line $l_i=l_j$ or   $t_i^{(l_i)}+t_j^{(l_j)}$
is the two point are on different lines $l_i\neq l_j$.

We find that
\begin{equation}
  \label{e:dP}
  \partial_{ij} P^{FT}_2(v_i-v_j)= (u_i-u_j)^T K^{-1} (v_i-v_j)+
  \begin{cases}
    \sg(t_i^{(l_i)}-t_j^{(l_j)})& \textrm{if}~l_i=l_j\cr
0&\textrm{otherwise}
  \end{cases}\,,
\end{equation}
and
\begin{equation}
  \label{e:ddP}
  \partial^2_{ij} P^{FT}_2(v_i-v_j)= (u_i-u_j)^T K^{-1} (u_i-u_j)+
  \begin{cases}
   2 \delta(t_i^{(l_i)}-t_j^{(l_j)})& \textrm{if}~l_i=l_j\cr
0&\textrm{otherwise}
  \end{cases}\,.
\end{equation}

We define the quantity

\begin{equation}
  \label{e:Q2}
  Q^{(2)}=\sum_{1\leq i<j\leq 4} k_i\cdot k_j\, P_2^{FT}(v_i-v_j)\,.
\end{equation}

 In this limit the expansion of CHL
model partition function
$\cZ_2^{(n_v)}$ is given by  in~\eqref{e:cl2limit}
where $O(q_i)$ do not contribute to the field    theory limit.
The  integration over the real part of the components
of the period matrix projects the integrand in the following way
\begin{equation}
\int_{-\frac12}^{\frac12}\,d^3\Ree\Omega\, \cZ_2^{(n_v)} F(\Omega,\bar \Omega)\,
= c_{n_v} F_0+ F_{123}+ a_{n_v}\,(F_{12}+F_{13}+F_{23})+b_{n_v}\,(F_1+F_2+F_3)\,,
\end{equation}
where $F(\Omega,\bar \Omega)= F_0+ \sum_{i=1}^3 \bar q_i F_i +
\sum_{1\leq i<j\leq 3} \bar q_i \bar q_j F_{ij}+\bar q_1\bar q_2\bar
q_3\, F_{123}+ c.c.+O(q_i \bar q_i)$  represents the
integrand of the two-loop amplitude.

When performing the field theory limit the integral takes the
form\footnote{A detailed analysis of these integrals will be given in~\cite{TourkineVanhove}.}

\begin{equation}
  \label{e:M42first}
  M^{(n_v)}_{4;2}= N_2\,t_8t_8R^4 \int_0^\infty {d^3L_i\over
  \Delta^{2+\epsilon}} \oint d^4t_i \, Y_S\, [W^{(2)}_1+ W^{(2)}_2]\, e^{Q^{(2)}}\,.
\end{equation}
The contribution of $W^{(2)}_1$ yields two kinds of two-loop double-box integrals
evaluated in $D=4-2\epsilon$; $I_{double-box}^{(D=4-2\epsilon)}[\ell^n]$  with $n=4,2,0$
powers of
loop momentum and  $s/u\, I_{double-box}^{(D=4-2\epsilon)}[\ell^m]$
with $m=2,0$ powers of loop momentum. Those integrals are multiplied by and overall
factor $s\times t_8t_8R^4$, $t\times t_8t_8R^4$ or $u\times t_8t_8R^4$ depending on the
channel according to the decomposition of $Y_S$ in~\eqref{e:Ysft}. 

The contribution of $W^{(2)}_2$ yields two-loop double-box integrals evaluated in
$D=6-2\epsilon$; $I_{double-box}^{(D=6-2\epsilon)}[\ell^n]$ with $n=2,0$ powers of
loop momentum multiplied by ${s\over u}\times t_8t_8R^4$ or
${t\over u}\times t_8t_8R^4$ or $t_8t_8R^4$ depending on the channel
according to  the decomposition of $Y_S$ in~\eqref{e:Ysft}.  
We therefore conclude that the field theory limit of the four-graviton two-loop amplitude
of the CHL models with various number of vector multiplets factorizes a $\partial^2 R^4$
term in four dimensions.

We remark that as in the one-loop case, the two-loop amplitude is free of ultraviolet
divergence, in agreement with the analysis of Grisaru~\cite{Grisaru:1976nn}.

%------------------------------------------------------------------------------------
\section{Non-renormalisation theorems}\label{sec:nonrenormalisation}

The analysis performed in this paper shows that the two-loop four-graviton amplitude in
$\cN=4$ pure supergravity factorizes a $\partial^2 R^4$ operator in each kinematical
sector. This result for the $R^4$ term
holds point wise in the moduli space of the string theory
amplitude. In the pure spinor formalism this is a direct consequence of
the fermionic zero mode saturation in the two-loop amplitude. At higher-loop since there will be
at least the same number of fermionic zero modes to saturate,  this implies that higher-loop four-graviton amplitudes will factorize  (at least) two
powers of external momenta on a $R^4$ term.\footnote{It is tempting to
  conjecture that the  higher-loop
  string amplitudes will have a form similar to the two-loop
  amplitude in~\eqref{e:twoloopstring} involving a  generalisation of
  $\mathcal Y_s$ in~\eqref{e:Ys}, maybe given by the ansatz 
  proposed in~\cite[eq.~(1.3)]{Matone:2005vm}.} This is in agreement
with the half-BPS nature of the $R^4$ term in $\cN=4$ models. We are then lead to the following
non-renormalisation theorem: the $R^4$ term will not receive any perturbative corrections
beyond one-loop in the four-graviton amplitudes.

Since the structure of the amplitude is the same in any dimension, a four-graviton
$L$-loop amplitude with $L\geq2$ in $D$ dimensions would have at worst the following
enhanced superficial ultraviolet behaviour $\Lambda^{(D-2)L-8}\, \partial^2R^4$ instead of
$\Lambda^{(D-2)L-6}\, R^4$, expected from supersymetry arguments \cite{Bossard:2011tq}.
This forbids the appearance of a three-loop ultraviolet divergence in four dimensions in
the four-graviton amplitude and delays it to four loops.

However, a fully supersymmetric $R^4$ three-loop ultraviolet counter-terms in four
dimensions has been constructed in~\cite{Bossard:2011tq}, so one can wonder why no
divergence occur. We provide here a few arguments that could explain why the $R^4$ term is
a protected operator in $\cN=4$ pure supergravity. 

It was argued in~\cite{Bachas:1996bp,Bachas:1997mc,Bachas:1997xn} that $R^4$ is a half-BPS
protected operator and does not receive perturbative corrections beyond one-loop in
heterotic string compactifications. These non-renormalisation theorems were confirmed
in~\cite{D'Hoker:2005ht} using the explicit evaluation of the genus-two four-graviton
heterotic amplitude derived in~\cite{D'Hoker:2001nj,D'Hoker:2002gw,D'Hoker:2005jc}. In
$D=4$ dimensions the CHL models with  $4\leq n_v\leq22$ vector multiplets obtained by an
asymmetric orbifold construction satisfy the same non-renormalisation theorems. For these
models the moduli space is $SU(1,1)/U(1)\times SO(6,n_v)/SO(6)\times SO(n_v)$. Since the
axion-dilaton  parametrizes the $SU(1,1)/U(1)$ factor it is natural to conjecture that
this moduli space will stay factorized  and that one can decouple the contributions from
the vector multiplets. If one can set to zero all the vector multiplets, this analysis
shows the existence of the $R^4$ non-renormalisation theorem in the pure $\cN=4$
supergravity case.

It was shown in~\cite{Bossard:2011tq} that the $SU(1,1)$-invariant
superspace volume vanishes and the $R^4$ super-invariant was
constructed as an
harmonic superspace integral over 3/4 of the full
superspace. 
 The structure of the amplitudes analyzed in this paper and
 the absence of three-loop divergence point to the fact
that this partial superspace integral is an F-term.
 
The existence of an off-shell formulation for $\cN=4$ conformal
 supergravity and linearized $\cN=4$ supergravity with six vector multiplets~\cite{Howe:1981gz,Bergshoeff:1980is,Howe:1982mt} makes this
F-term nature plausible in the Poincar\'e pure
supergravity.

What makes the $\cN=4$ supergravity case special compared to the other
$5\leq\cN\leq8$ cases is the anomalous $U(1)$
symmetry~\cite{Marcus:1985yy}. Therefore even without the existence of
an off-shell formalism,  this anomaly could 
make the $R^4$ term special
and be the reason why it turns out to be ruled out as a possible
counter-term in four-graviton amplitude in four dimensions.
Because of the $U(1)$-anomaly, full superspace
integrals of functions of the axion-dilaton
superfield $\mathbb S=S+\cdots$ are allowed~\cite{Bossard:2011tq}
\begin{equation}
I=\kappa_{(4)}^4\,\int d^4x
d^{16}\theta\, E(x,\theta)\,F(\mathbb S)= \kappa_{(4)}^4\,\int d^4x\,
\sqrt{-g}\, f(S)\,R^4+\textrm{susy~completion}\,,
\end{equation}
suggesting a three-loop divergence in the higher-point field theory amplitudes with
four gravitons and scalar fields. Since one can write full superspace
for $\partial^2 R^4$ in terms of the gravitino
$\int d^{16}\theta\, E(x,\theta) (\chi\bar\chi)^2$, one should expect a four-loop divergence
in the four-graviton amplitude in four dimensions.

%------------------------------------------------------------------------------------
\section*{Acknowledgements}

We would like to thank C. Bachas, G. Bossard, E. D'Hoker, P.S. Howe, J. Russo,
A. Sen, and E. Sokatchev  for discussions, and Mike Duff and Kelly
Stelle for discussions about $\cN=4$ supergravity.
PV would like to thank the Newton Institute for the hospitality when
this work was carried out.

%------------------------------------------------------------------------------------


\begin{thebibliography}{99}




%\cite{de Roo:1984gd}
\bibitem{de Roo:1984gd}
M.~de Roo,
``Matter Coupling in ${\mathcal{N}}\!=4$ Supergravity,''
Nucl.\ Phys.\ B {\bf 255} (1985) 515.
%%CITATION = NUPHA,B255,515;%%



%\cite{Chaudhuri:1995fk}
\bibitem{Chaudhuri:1995fk}
S.~Chaudhuri, G.~Hockney and J.~D.~Lykken,
``Maximally Supersymmetric String Theories in $D < 10$,''
Phys.\ Rev.\ Lett.\ {\bf 75} (1995) 2264
[arXiv:hep-th/9505054].
%%CITATION = PRLTA,75,2264;%%


%\cite{Chaudhuri:1995bf}
\bibitem{Chaudhuri:1995bf}
S.~Chaudhuri and J.~Polchinski,
``Moduli Space of Chl Strings,''
Phys.\ Rev.\ D {\bf 52} (1995) 7168
[arXiv:hep-th/9506048].
%%CITATION = PHRVA,D52,7168;%%



%\cite{Chaudhuri:1995dj}
\bibitem{Chaudhuri:1995dj}
S.~Chaudhuri and D.~A.~Lowe,
``Type IIA Heterotic Duals with Maximal Supersymmetry,''
Nucl.\ Phys.\ B {\bf 459} (1996) 113
[arXiv:hep-th/9508144].
%%CITATION = NUPHA,B459,113;%%


%\cite{Schwarz:1995bj}
\bibitem{Schwarz:1995bj}
J.~H.~Schwarz and A.~Sen,
``Type Iia Dual of the Six-Dimensional Chl Compactification,''
Phys.\ Lett.\ B {\bf 357} (1995) 323
[arXiv:hep-th/9507027].
%%CITATION = PHLTA,B357,323;%%



%\cite{Aspinwall:1995fw}
\bibitem{Aspinwall:1995fw}
P.~S.~Aspinwall,
``Some Relationships Between Dualities in String Theory,''
Nucl.\ Phys.\ Proc.\ Suppl.\ {\bf 46} (1996) 30
[arXiv:hep-th/9508154].
%%CITATION = NUPHZ,46,30;%%




%\cite{Bachas:1996bp}
\bibitem{Bachas:1996bp}
C.~Bachas and E.~Kiritsis,
``$ F^4$ Terms in ${\mathcal{N}}\!=4$ String Vacua,''
Nucl.\ Phys.\ Proc.\ Suppl.\ {\bf 55B} (1997) 194
[arXiv:hep-th/9611205].
%%CITATION = NUPHZ,55B,194;%%



%\cite{Bachas:1997mc}
\bibitem{Bachas:1997mc}
C.~Bachas, C.~Fabre, E.~Kiritsis, N.~A.~Obers and P.~Vanhove,
``Heterotic/Type-I Duality and D-Brane Instantons,''
Nucl.\ Phys.\ B {\bf 509} (1998) 33
[arXiv:hep-th/9707126].
%%CITATION = NUPHA,B509,33;%%


%\cite{Bachas:1997xn}
\bibitem{Bachas:1997xn}
C.~Bachas,
``Heterotic Versus Type I,''
Nucl.\ Phys.\ Proc.\ Suppl.\ {\bf 68} (1998) 348
[arXiv:hep-th/9710102].
%%CITATION = NUPHZ,68,348;%%

%\cite{D'Hoker:2005ht}
\bibitem{D'Hoker:2005ht}
E.~D'Hoker, M.~Gutperle and D.~H.~Phong,
``Two-Loop Superstrings and S-Duality,''
Nucl.\ Phys.\ B {\bf 722} (2005) 81
[arXiv:hep-th/0503180].
%%CITATION = NUPHA,B722,81;%%



%\cite{D'Hoker:2001nj}
\bibitem{D'Hoker:2001nj}
E.~D'Hoker and D.~H.~Phong,
``Two-Loop Superstrings I, Main Formulas,''
Phys.\ Lett.\ B {\bf 529} (2002) 241
[arXiv:hep-th/0110247].
%%CITATION = PHLTA,B529,241;%%



%\cite{D'Hoker:2002gw}
\bibitem{D'Hoker:2002gw}
E.~D'Hoker and D.~H.~Phong,
``Lectures on Two-Loop Superstrings,''
arXiv:hep-th/0211111.
%%CITATION = HEP-TH/0211111;%%


%\cite{D'Hoker:2005jc}
\bibitem{D'Hoker:2005jc}
E.~D'Hoker and D.~H.~Phong,
``Two-Loop Superstrings Vi: Non-Renormalization Theorems and the 4-Point Function,''
Nucl.\ Phys.\ B {\bf 715} (2005) 3
[arXiv:hep-th/0501197].
%%CITATION = NUPHA,B715,3;%%


\bibitem{Bern3loopN4} Zvi Bern, Scott Davies, Tristan Dennen, Yu-tin
  Huang ``Absence of Three-Loop Four-Point Divergences in N=4
  Supergravity'', [arXiv:1202.3423]

%\cite{Sakai:1986bi}
\bibitem{Sakai:1986bi}
N.~Sakai and Y.~Tanii,
``One Loop Amplitudes and Effective Action in Superstring Theories,''
Nucl.\ Phys.\ B {\bf 287} (1987) 457.
%%CITATION = NUPHA,B287,457;%%


%\cite{Green:1999pv}
\bibitem{Green:1999pv}
M.~B.~Green and P.~Vanhove,
``The Low Energy Expansion of the One-Loop Type II Superstring Amplitude,''
Phys.\ Rev.\ D {\bf 61} (2000) 104011
[arXiv:hep-th/9910056].
%%CITATION = PHRVA,D61,104011;%%


%\cite{Green:2008uj}
\bibitem{Green:2008uj}
M.~B.~Green, J.~G.~Russo and P.~Vanhove,
``Low Energy Expansion of the Four-Particle Genus-One Amplitude in Type II Superstring Theory,''
JHEP {\bf 0802} (2008) 020
[arXiv:0801.0322 [hep-th]].
%%CITATION = JHEPA,0802,020;%%







%\cite{Berkovits:2005df}
\bibitem{Berkovits:2005df}
N.~Berkovits,
``Super-Poincare Covariant Two-Loop Superstring Amplitudes,''
JHEP {\bf 0601} (2006) 005
[arXiv:hep-th/0503197].
%%CITATION = JHEPA,0601,005;%%


%\cite{Berkovits:2005ng}
\bibitem{Berkovits:2005ng}
N.~Berkovits and C.~R.~Mafra,
``Equivalence of Two-Loop Superstring Amplitudes in the Pure Spinor and RNS Formalisms,''
Phys.\ Rev.\ Lett.\ {\bf 96} (2006) 011602
[arXiv:hep-th/0509234].
%%CITATION = PRLTA,96,011602;%%


\bibitem{TourkineVanhove} P. Tourkine and P. Vanhove, ``Four-graviton
  amplitudes in $\cN=4$ supergravity models'', to appear


%\cite{Green:1982sw}
\bibitem{Green:1982sw}
M.~B.~Green, J.~H.~Schwarz and L.~Brink,
``${\mathcal{N}}\!=4$ Yang-Mills and ${\mathcal{N}}\!=8$ Supergravity as Limits of String Theories,''
Nucl.\ Phys.\ B {\bf 198} (1982) 474.
%%CITATION = NUPHA,B198,474;%%


%\cite{BjerrumBohr:2008vc}
\bibitem{BjerrumBohr:2008vc}
N.~E.~J.~Bjerrum-Bohr and P.~Vanhove,
``Explicit Cancellation of Triangles in One-Loop Gravity Amplitudes,''
JHEP {\bf 0804} (2008) 065
[arXiv:0802.0868 [hep-th]].
%%CITATION = JHEPA,0804,065;%%



%\cite{BjerrumBohr:2008ji}
\bibitem{BjerrumBohr:2008ji}
N.~E.~J.~Bjerrum-Bohr and P.~Vanhove,
``Absence of Triangles in Maximal Supergravity Amplitudes,''
JHEP {\bf 0810} (2008) 006
[arXiv:0805.3682 [hep-th]].
%%CITATION = JHEPA,0810,006;%%




%\cite{Dunbar:2010fy}
\bibitem{Dunbar:2010fy}
D.~C.~Dunbar, J.~H.~Ettle and W.~B.~Perkins,
``Perturbative Expansion of N<8 Supergravity,''
Phys.\ Rev.\ D {\bf 83} (2011) 065015
[arXiv:1011.5378 [hep-th]].
%%CITATION = PHRVA,D83,065015;%%


%\cite{Dunbar:2011xw}
\bibitem{Dunbar:2011xw}
D.~C.~Dunbar, J.~H.~Ettle and W.~B.~Perkins,
``Obtaining One-Loop Gravity Amplitudes Using Spurious Singularities,''
Phys.\ Rev.\ D {\bf 84} (2011) 125029
[arXiv:1109.4827 [hep-th]].
%%CITATION = PHRVA,D84,125029;%%


%\cite{Dunbar:2011dw}
\bibitem{Dunbar:2011dw}
D.~C.~Dunbar, J.~H.~Ettle and W.~B.~Perkins,
``The N-Point Mhv One-Loop Amplitude in ${\mathcal{N}}\!=4$ Supergravity,''
arXiv:1111.1153 [hep-th].
%%CITATION = ARXIV:1111.1153;%%


%\cite{Bern:2011rj}
\bibitem{Bern:2011rj}
Z.~Bern, C.~Boucher-Veronneau and H.~Johansson,
``$N\geq 4$ Supergravity Amplitudes from Gauge Theory at One Loop,''
Phys.\ Rev.\ D {\bf 84} (2011) 105035
[arXiv:1107.1935 [hep-th]].
%%CITATION = PHRVA,D84,105035;%%


%\cite{Fischler:1979yk}
\bibitem{Fischler:1979yk}
M.~Fischler,
``Finiteness Calculations for $O(4)$ Through $O(8)$ Extended Supergravity and $O(4)$ Supergravity Coupled to Selfdual $O(4)$ Matter,''
Phys.\ Rev.\ D {\bf 20} (1979) 396.
%%CITATION = PHRVA,D20,396;%%





%\cite{Green:2008bf}
\bibitem{Green:2008bf}
M.~B.~Green, J.~G.~Russo and P.~Vanhove,
``Modular Properties of Two-Loop Maximal Supergravity and Connections with String Theory,''
JHEP {\bf 0807} (2008) 126
[arXiv:0807.0389 [hep-th]].
%%CITATION = JHEPA,0807,126;%%



%\cite{Grisaru:1976nn}
\bibitem{Grisaru:1976nn}
  M.~T.~Grisaru,
  ``Two Loop Renormalizability of Supergravity,''
  Phys.\ Lett.\ B {\bf 66} (1977) 75.
  %%CITATION = PHLTA,B66,75;%%

\bibitem{Matone:2005vm}
  M.~Matone and R.~Volpato,
  ``Higher genus superstring amplitudes from the geometry of moduli space,''
  Nucl.\ Phys.\ B {\bf 732} (2006) 321
  [hep-th/0506231].
  %%CITATION = HEP-TH/0506231;%%


%\cite{Bossard:2011tq}
\bibitem{Bossard:2011tq}
G.~Bossard, P.~S.~Howe, K.~S.~Stelle and P.~Vanhove,
``The Vanishing Volume of $d=4$ Superspace,''
Class.\ Quant.\ Grav.\ {\bf 28} (2011) 215005
[arXiv:1105.6087 [hep-th]].
%%CITATION = CQGRD,28,215005;%%


%\cite{Howe:1981gz}
\bibitem{Howe:1981gz}
P.~S.~Howe,
``Supergravity in Superspace,''
Nucl.\ Phys.\ B {\bf 199} (1982) 309.
%%CITATION = NUPHA,B199,309;%%

\bibitem{Bergshoeff:1980is}
E.~Bergshoeff, M.~de Roo and B.~de Wit,
``Extended Conformal Supergravity,''
Nucl.\ Phys.\ B {\bf 182} (1981) 173.
%%CITATION = NUPHA,B182,173;%%


\bibitem{Howe:1982mt}
P.~S.~Howe, H.~Nicolai and A.~Van Proeyen,
``Auxiliary Fields and a Superspace Lagrangian for Linearized Ten-Dimensional Supergravity,''
Phys.\ Lett.\ B {\bf 112} (1982) 446.
%%CITATION = PHLTA,B112,446;%%




%\cite{Marcus:1985yy}
\bibitem{Marcus:1985yy}
N.~Marcus,
``Composite Anomalies in Supergravity,''
Phys.\ Lett.\ B {\bf 157} (1985) 383.
%%CITATION = PHLTA,B157,383;%%





\end{thebibliography}
\end{document}